\documentclass[preprint, 12pt, a4paper]{revtex4}

\usepackage{ulem}
\usepackage[utf8]{inputenc}
\usepackage[T1]{fontenc}
\usepackage{amsmath}
\usepackage{amsfonts}
\usepackage{amssymb}
\usepackage{graphics,graphicx}
\usepackage[unicode=true,bookmarks=false,breaklinks=false,pdfborder={0 0 1},colorlinks=true]
 {hyperref}
\hypersetup{
 citecolor=blue,linkcolor=blue,urlcolor=blue}

\usepackage{graphicx,color}
\usepackage{subfigure} 
\usepackage{wrapfig} 
\usepackage{epstopdf}
\graphicspath{{figuras/}}

\begin{document}

\title{AdS/QCD oddball masses and Odderon Regge trajectory from a  twist-five operator approach}
\author{Eduardo Folco Capossoli$^{1}$}
\email[Eletronic address: ]{eduardo\_capossoli@cp2.g12.br}
\author{J. P. Morais Graça$^{2}$}
\email[Eletronic address: ]{jpmorais@gmail.com}
\author{Henrique Boschi-Filho$^{2}$}
\email[Eletronic address: ]{hboschi@gmail.com}  
\affiliation{$^1$Departamento de F\'{\i}sica and Mestrado Profissional em Práticas de Educa\c{c}\~{a}o B\'{a}sica (MPPEB),  Col\'egio Pedro II, 20.921-903 - Rio de Janeiro-RJ - Brazil \\ 
 $^2$Instituto de F\'{\i}sica, Universidade Federal do Rio de Janeiro, 21.941-972 - Rio de Janeiro-RJ - Brazil}

\begin{abstract}
In this work, we consider a massive gauge boson field in AdS$_5$ dual to odd glueball states with twist-5 operator in 4D Minkowski spacetime. Introducing an IR cutoff we break the conformal symmetry of the boundary theory allowing us to calculate the glueball masses with odd spins using Dirichlet and Neumann boundary conditions. Then, from these masses we construct the corresponding Regge trajectories associated with the odderon. Our results are compatible with the ones in the literature. 
\end{abstract}

\maketitle

\section{Introduction}

The history of what would come to be called glueballs goes back to the early days of hadronic physics, before the emergence of QCD. At that time, hadron-hadron scattering processes at high-energies and low transferred momenta,  written in terms of the Mandelstan variables, $s \gg m^2 \simeq -t $, 
were ruled by Regge theory. In that  scenario, Regge proposed that in the hadronic processes ``particles were exchanged'', for instance, as a meson ($\rho$, $\omega$, etc.) or as a ``Reggeons''. In both cases,  their scattering amplitudes, in the $t$ channel, behaves like ${\cal A}(s,t) \sim s^{\alpha(t)}$. If one considers a family or a set of resonances sharing the same quantum numbers, one can display them in a plane $[ t\equiv m^2, \alpha(t)\equiv J]$  fulfilling a linear relationship written as
\begin{equation}\label{Regge}
    J(m^2) = \alpha' m^2 + \alpha_0\,,
\end{equation}
\noindent where $J$ is total angular momentum, $m$ is the mass of the Reggeized particle,  $\alpha'$  and $\alpha_0$ are two constants. The above relationship plotted in a Chew-Frautschi plane is known as the Regge trajectory. 

If these Reggeized particles are Reggeized gluons, one has the so-called glueballs. Glueballs are represented by their total angular momentum $J$ and their vacuum quantum numbers $P$, $C$ and $I$, where $P$ is the $P-$parity (or spatial inversion), $C$ is the $C-$parity (or charge  conjugation) and $I$ is the isospin. By using the spectroscopy notation one has $J^{PC}$, omitting the isospin $I$ since it is zero for all states considered here. For a review on glueballs, one can see Ref. \cite{Mathieu:2008me}.

From now on, let us focus on oddballs or glueballs with odd angular momentum ($J\geq 1$),  and  quantum numbers taken as $P=-1$, $C=-1$, and  $I=0$, such as, $1^{--}, 3^{--}, 5^{--}, \cdots$. Odd spin glueballs are particularly interesting because they lie on the Regge trajectory of an exchanged Reggeon called odderon.

\begin{figure}[ht]
	\centering
	\includegraphics[scale = 0.6]{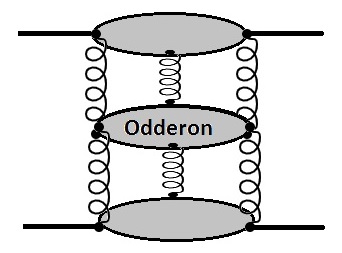}
	\caption{The odderon as colorless $C$-odd three-gluon bound state exchanged in a hadron-hadron scattering.}
	\label{3gluons}
\end{figure}

 In the context of perturbative QCD, the odderon is described by the Bartels-Kwiecinski-Praszalowicz (BKP) equation \cite{Bartels:1978fc, Bartels:1980pe, Kwiecinski:1980wb}, as a colorless $C$-odd three reggeons (gluons) compound state in the $t$ channel, as can be seen  pictorially in Fig. \ref{3gluons}.  An interesting review on the odderon physics can be seen in Ref. \cite{Ewerz:2003xi}.

The original proposal for the existence of the odderon in early 1970s appeared in Ref. \cite{Lukaszuk:1973nt}, the first attempts for its measurement in  Refs. \cite{Hill:1973bq, 
Bonamy:1973dz}, and continued through the decades of 1980 and 1990 \cite{Apokin:1982kw, Augier:1993sz}. Note that all these collaborations did not provide reliable experimental evidence about the existence of the odderon. Recently, the outstanding efforts done in TOTEM and D0 Collaborations, analyzing the cross sections for $pp$ and $p \bar{p}$, and eventually their differences, $\Delta \sigma (s) = \sigma^{pp}(s) - \sigma^{p{\bar p}}(s) \propto \ln s$, supported the existence of the odderon with $3.4 \sigma$ of significance \cite{TOTEM:2017sdy}. In Ref. \cite{TOTEM:2020zzr} the significance was improved to $5.2\sigma - 5.7 \sigma$. The combination of these results may be considered sufficient to give the odderon experimentally discovered.

Motivated by this recent discovery, in the present work we are interested in odd spin glueballs $J^{--}$, with ($J\geq 1$). Our aim here is to contribute with new insights and proposals to compute the oddball masses and then calculate the corresponding Regge trajectory related to the odderon. To do so, we will resort to an AdS/QCD model inspired by a  duality proposed by Maldacena \cite{Maldacena:1997re}. AdS/QCD is a suitable approach to deal with QCD phenomenology in the nonperturbative regime,  where glueballs are formed. The AdS/QCD model used here is known as the hardwall model, as proposed independently in Refs. \cite{Polchinski:2001tt, BoschiFilho:2002vd, BoschiFilho:2002ta}. 
In this model, conformal symmetry is broken due to the introduction of an IR cutoff $z_{\rm max}$ in the holographic coordinate $z$ and  considering a slice of the anti–de Sitter (AdS) space, given by the interval $0 \leq z \leq z_{\rm max}$. In Ref. \cite{Erlich:2005qh} the authors used the hardwall model to compute the masses of vector mesons. In the last 20 years the AdS/QCD community have done many contributions offering many approaches to deal with glueballs and correlated issues. Here, one can see in Refs. \cite{BoschiFilho:2005yh, Colangelo:2007pt,
  Wang:2009wx, Huang:2007fv, Afonin:2012jn, BoschiFilho:2012xr, Capossoli:2013kb, Li:2013oda, Capossoli:2015ywa, Brunner:2015oqa, Brunner:2015yha, Capossoli:2016kcr, Chen:2015zhh, Capossoli:2016ydo, Brunner:2016ygk, Rodrigues:2016cdb, FolcoCapossoli:2016ejd, Rinaldi:2017wdn, Afonin:2018era, Rinaldi:2018yhf, FolcoCapossoli:2019imm, Rinaldi:2020ssz, Rinaldi:2021dxh, Zhang:2021itx} an incomplete list of those contributions which take into account even and odd spin glueballs, top-down and botton-up holographic models, considering anomalous dimension, dynamical AdS/QCD models, deformed AdS metric space, Einstein-Maxwell-dilaton background, among other proposals.
  
  This work is organized as follows: in section \ref{v1} we present our holographic description of the odd spin glueballs with $J^{PC}$=$1^{--}$, $3^{--}$, $5^{--}$, etc, starting from a twist five operator in a massive vector gauge boson. In section \ref{res}, we calculate oddball masses using Dirichlet and Neumann boundary conditions and construct some proposals to the odderon Regge trajectory. In this section we also compare our results for masses and trajectories with known results from the literature. Finally, in section \ref{conc} we present our last comments, interpretations and conclusions. 
  

\section{Holographic description of Odd spin glueballs}\label{v1}

Here, in this section, we are going to present the description of a vector glueball state within the AdS/QCD model, compute the masses for $J^{PC}$=$1^{--}$, $3^{--}$, $5^{--}$, etc,  and construct the Regge trajectory associated with the odderon.

First of all, let us emphasize the main feature of this work. As the ground state for the odd spin glueballs, $1^{--}$ is a vector object, living at the UV boundary, we start our calculation within the holographic hardwall model by relating it to a five-dimensional massive gauge boson field defined in the AdS$_5$ space. This procedure, which relates operators in the four-dimensional theory to  fields in the bulk of five-dimensional space, represents the accomplishment of the AdS/CFT correspondence. 

 The twist or twist dimension, represented by $\tau$, is given by the conformal dimension ($\Delta$) of an operator minus its spin. In particular it will be shown that the conformal dimension of the state $1^{--}$ is $\Delta = 6$, and then $\tau = \Delta - J =5$. In this sense we are going to refer to our model as a twist-five approach.

Note that in the Ref. \cite{Capossoli:2013kb} the authors dealt with oddballs and odderon Regge trajectories, also using the hardwall model, however relating the ground state for the odd glueballs $1^{--}$ to a massive scalar field in the AdS$_5$ space. In the reference \cite{Chen:2015zhh} the authors also started their computation, within the hardwall model, from a massive boson field in AdS side. However, among many exotic glueball states, the authors considered only one odd glueball state, namely the state $1^{--}$.

Now let us introduce the action for a five-dimensional massive gauge boson field $A_m$ which represents the physical vector glueball at four-dimensional boundary theory, so that:
\begin{equation}\label{vec_hw}
S = -\frac{1}{2}\int d^5 x \sqrt{-g}\; [ \frac{1}{2} g^{pm} g^{qn} F_{mn} F_{pq} +M_5^2 g^{pm} A_p A_m]\,.
\end{equation} 
\noindent Note that vector field stress tensor is assumed as $F_{mn} = \partial_m A_n - \partial_n A_m$ and $M_5$ is the mass of the gauge boson field. Besides $g$ is the determinant of the metric $g_{mn}$ of the $AdS_5$ space, given by:
\begin{equation}\label{metric} 
ds^2 = g_{mn} dx^m dx^n= \frac{L^2}{z^2} \, (dz^2 + \eta_{\mu \nu}dy^\mu dy^\nu)\,,
\end{equation}
\noindent where $z$ is the holographic coordinate, and $L$ is the AdS radius. From now on we take $L=1$ throughout the text and $\eta_{\mu \nu}$ with signature $(-,+,+,+)$ is the Minkowski flat spacetime metric.

By computing  $\delta S / \delta A_n = 0$, one obtains the corresponding equations of motion:
\begin{equation}\label{f16}
\partial_p [ \sqrt{-g} g^{m p} g^{nq}\; F_{mn}] - M_5^2 \sqrt{-g} g^{nq} A_n = 0\,. 
\end{equation}
Plugging the AdS metric in the above equation and considering $p=z,\mu$, one finds:
\begin{equation}\label{eom_vec_4}
\partial_z \left[\left(\frac{1}{z}\right) F_{zn} \eta^{nq}\right] + \partial_{\mu}\left[\left(\frac{1}{z}\right) \eta^{m \mu} F_{m n} \eta^{nq}\right] -  M_5^2 \left(\frac{1}{z}\right)^3 A_n \;\eta^{nq}= 0,
\end{equation}
\noindent with $g^{mn} = z^{2} \eta^{mn}$. 

In order to solve the above equation, firstly, we will use an ansatz for a plane wave with four-momentum $q_{\mu}$, which is propagating in the in the transverse coordinates $x^{\mu}$, given by:
\begin{equation}\label{anvec}
A_{\rho} (z, x^{\mu}) = \epsilon_{\rho} \, v(z) e^{i q_{\mu} x^{\mu}} , 
\end{equation}
\noindent where $\epsilon_{\rho}$  is the polarization four-vector defined in the transverse space to $z$ coordinate, and the plane wave amplitude depends on of $z$ coordinate, only. Note that $\epsilon^{\rho} \epsilon_{\rho} = \eta^{\rho \lambda} \epsilon_{\rho} \epsilon_{\lambda}=1$.

Following Ref. \cite{Erlich:2005qh} we are going to consider $A_z=0$, and then, it implies that $F_{zn}=\partial_{z} A_n$. Besides we  choose $\partial_{\mu} A^{\mu}=0$, which implies $q^{\mu} \epsilon_{\mu} = \eta^{\mu \lambda} q^{\mu} \epsilon_{\lambda} = q \cdot \epsilon = 0$ ensuring that the field can be written as a plane wave. Therefore, one can get
\begin{eqnarray} \label{fields}
\eta^{m \mu} \partial_{\mu}  F_{m n} & = &  \eta^{m \mu} (i q_{\mu}) (i q_m A_n - \partial_m A_n) \nonumber \\
& = & -q^2 A_n - (\partial_{\mu} A^{\mu}) \nonumber \\
& = & -q^2 A_n
\end{eqnarray}

At this point, we can rewrite Eq. \eqref{eom_vec_4} as
\begin{equation}
\partial_z \left[\left(\frac{1}{z}\right) \partial_z A_{n} \eta^{nq}\right] - \left(\frac{1}{z}\right) q^2 A_n \eta^{nq} -  M_5^2 \left(\frac{1}{z}\right)^3 A_n \;\eta^{nq}= 0, 
\end{equation}
or using \eqref{anvec}, one has:
\begin{equation}\label{eom_vec_5}
\left\{ \partial_z \left[\left(\frac{1}{z}\right) \partial_z v(z)\right] - \left(\frac{1}{z}\right) q^2 v(z) -  M_5^2 \left(\frac{1}{z}\right)^3 v(z) \right\} \cdot e^{i q_{\mu} x^{\mu}} \epsilon^q = 0\,.
\end{equation}

Defining $v(z) = z\,\psi(z)$ and plugging it in above equation, so that:
\begin{eqnarray}
z^2 \frac{d^2 \psi(z)}{dz^2} + z \frac{d \psi(z)}{dz} - [(1 + M_5^2) + q^2 z^2] = 0\,,
\end{eqnarray}
whose solutions are given by a linear combination of Bessel ($J_{\nu}$) and Neumann ($Y_{\nu}$) functions: 
\begin{equation}\label{psi}
    \psi (z) = {\cal A}_{\nu, k} J_{\nu}(m_{\nu, k} \,z)+ {\cal B}_{\nu, k} Y_{\nu}(m_{\nu, k} \,z)\,,
\end{equation}
where ${\cal A}_{\nu, k}$ and ${\cal B}_{\nu, k}$ are normalization constants, the index $\nu = \sqrt{M_5^2 +1}$ and $m_{\nu, k}^2 = -q^2$ will be the mass squared of the odd spin glueballs at the boundary. Note that $k=1,2,3, \cdots$ denote radial excitations, with $k= 1$ for the ground state. As we are interested in regular solutions inside the bulk, we are just considering the Bessel solution and disregarding the Neumann one. Now, by plugging Eq. \eqref{psi}
 in Eq. \eqref{anvec} we can construct the complete solution for the field $A_{\rho} (z, x^{\mu})$, so that:
\begin{equation}\label{anveccomp}
A_{\rho} (z, x^{\mu}) = {\cal A}_{\nu, k}\, z \, J_{\nu}(m_{\nu, k} \,z) e^{i q_{\mu} x^{\mu}} \epsilon_{\rho}\,.
\end{equation}
In order to get the odd spin glueball masses we are going to impose boundary conditions, such as Dirichlet and Neumann, on the vector field $A_{\rho} (z, x^{\mu})$. Before we impose those boundary conditions, one has to resort to the AdS/CFT dictionary and learn how to relate the gauge boson bulk mass ($M_5$) and the conformal dimension ($\Delta$) of the corresponding operator (${\cal O}$) in the four-dimensional theory. Such a relationship is written as:
\begin{equation}\label{brod}
M^2_5 = (\Delta - p) (\Delta + p - 4), 
\end{equation}
where, $p$ represents the $p-$form index. Here we will consider $p=1$.

In particular, for the glueball ground state $1^{--}$, it is associated to an operator ${\cal O}_6$ at the UV, given by \cite{Wang:2009wx, Capossoli:2013kb, Csaki:1998qr, Brower:2000rp}:
\begin{equation}
 {\cal O}_{6} =\text{Sym}\,\text{Tr}\left( {\tilde{F}_{\mu \nu}}F^2\right)\,.
 \end{equation} 
From this operator one can infer that the scaling or conformal  dimension should be $\Delta = 6$. As a consequence the ground state for oddballs is associated with a twist-five operator, since the twist $\tau$ is defined as the dimension minus spin, and then, $\tau = \Delta - J = 5$.

To construct higher spin glueball states we will follow Ref. \cite{deTeramond:2005su} where the authors proposed to raise the total angular momentum by inserting  symmetrised covariant derivatives in a given operator with spin $S$. After this insertion one gets
\begin{equation}\label{6+J}
{\cal O}_{6 + \ell} = SymTr\left( {\tilde{F}_{\mu \nu}}F D_{\lbrace\mu1 \cdots} D_{\mu \ell \rbrace}F\right),
\end{equation}

\noindent with conformal dimension $\Delta = 6 + \ell$ and total angular momentum $J= 1+\ell$. So to obtain the states $3^{--}, 5^ {--},$ etc, we take $\ell=2, 4, \cdots$. Then, all odd spin states in this formulation will have twist $\tau=5$.

Now, replacing $\Delta = 6 + \ell$ in Eq. \eqref{brod}, one has 
\begin{equation}\label{hsi}
M^2_{5} = (\Delta +\ell - p) (\Delta + \ell + p - 4)\,; \qquad ({\rm even}\, \ell \geq 2, \,\,p=1).
\end{equation}
In this work we consider all odd spin glueball states associated with $p=1$ forms. 

\section{Results achieved}\label{res}

In this section we will present our results for the masses of higher odd spin glueballs as well as the Regge trajectories associated with the odderon  achieved from our holographic hardwall model within a twist five operator approach, considering the usual Dirichlet and Neumann boundary conditions. In order to compare our results for oddball masses, we consider as benchmarks other results found within different approaches. Those data were extracted from the literature and are summarized in Table \ref{oddlit}. Note that there are no experimental data for glueball masses and there are  
few values from lattice simulations, QCD rules, Wilson loops, and semirelativistic potentials. In particular, lattice simulations require strong computational efforts to compute high spin glueball masses.

Regarding the odderon's Regge trajetory, one should note that the precise values for its slope  ($\alpha'$) and intercept ($\alpha_0$), are not consensus and are still open questions. Almost twenty years ago the Ref. \cite{Kovchegov:2003dm} pointed out that Refs. \cite{Janik:1998xj, Korchemsky:2001nx, Bartels:1999yt}, considering different solutions for BKP equation, found divergent values for the odderon's intercept. Besides, in Ref. \cite{Bartels:1999yt} one can see the largest intercept reported which is close to the unity.

In particular, two different odderon's Regge trajectories were proposed in Ref. \cite{LlanesEstrada:2005jf}, which are
\begin{equation}\label{l1}
J^{RMB}(m^2) = 0.23 m^2 -0.88\,,
\end{equation}
\noindent obtained by using a relativistic many-body (RMB) model, and 
\begin{equation}\label{l2}
J^{NRCM}(m^2) = 0.18 m^2 +0.25\,,  
\end{equation}
\noindent based on a nonrelativistic constituent model (NRCM).

\begin{table}
{\small 
\centering
\begin{tabular}{|c|c|c|c|c|c|c|}
\hline
Models used &  \multicolumn{6}{c|}{{Odd  Glueball States $J^{PC}$}}  \\  
\cline{2-7}
 &\qquad  $1^{ - - }$ \qquad & \qquad  $3^{- - }$ \qquad & \qquad  $5^{- - }$ \qquad & \qquad  $7^{- - }$  \qquad & \qquad $9^{- - }$ \qquad & \qquad $11^{- - }$ \qquad \\
\hline \hline
 $SU(3)$ gauge th.  \cite{Athenodorou:2020ani}             
& 4.03(7)   &   &   &   &  &    \\ \hline
{ Iso. lattice} \cite{Meyer:2004jc, Meyer:2004gx}            & { 3.240(330)(150)} & {4.330(260)(200)} &      &    &&           \\ \hline
{Anis. lattice} \cite{Chen:2005mg}                         & {3.830(40)(190)} & {4.200(45)(200)} &      &    &&            \\ \hline
{Anis. lattice} \cite{Morningstar:1999rf}                         & {3.850 (50) (190)} & {4.130 (90) (200)} &      &    &&           \\ \hline
{QCD sum rules} \cite{Chen:2021cjr}                         & {$3.29^{+1.49}_{-0.32}$} & {$3.47^{+?}_{-0.50}$} &      &   &&            \\ \hline
{Doub. pole model} \cite{Szanyi:2019kkn}                         & {$3.001$} & {$4.416$} & 5.498     &         &&      \\ \hline

 Relat. many body \cite{LlanesEstrada:2005jf}             &  3.95 &  4.15 &  5.05 &  5.90   &&    \\ \hline
 Nonrelat. const.  \cite{LlanesEstrada:2005jf}       
&  3.49 &  3.92 &  5.15 &  6.14     &&  \\ \hline
 Wilson loop \cite{Kaidalov:1999yd}                             
&  3.49 &  4.03 &      &     &&          \\ \hline
Vac. correlator \cite{Kaidalov:2005kz}                       
&  3.02 &  3.49 &  4.18 & 4.96   &&       \\ \hline
 Vac. correlator \cite{Kaidalov:2005kz}                       
& { 3.32} &  3.83 & { 4.59} & {5.25}     &&    \\ \hline
 Semirelat. pot.   \cite{Mathieu:2008pb}               
& {3.99} & {4.16} & {5.26} &   &&           \\ \hline
Hardwall twist 4 D  \cite{Capossoli:2013kb}
& 3.24 & 4.09 & 4.93 & 5.75 & 6.57 & 7.38  \\ \hline
Hardwall twist 4 N \cite{Capossoli:2013kb}
& 3.24 & 4.21 & 5.17 & 6.13 & 7.09 & 8.04
 \\ \hline
Modified softwall \cite{Capossoli:2015ywa}
& 2.82 & 3.94 & 5.03 & 6.11 & 7.19 & 8.26  \\ \hline 
\end{tabular}}
\caption{{Glueball masses for $J^{PC}$ states expressed in GeV, with odd $J$, achieved with nonholographic and some holographic models from the literature.Note the abbreviations in the first column of this table can be read as: $SU(3)$
gauge th. ($SU(3)$ gauge theory in (3+1)d); Iso. lattice (Isotropic lattice); Anis. lattice (Anisotropic lattice); Doub.pole model (Double pole model), Relat. many body (Relativistic many body); Nonrelat. const. (Nonrelativistic constituent); Vac. correlator (Vacuum correlator) and Semirelat. pot. (Semirelativistic potential).}}
\label{oddlit}
\end{table}

\subsubsection{Dirichlet boundary condition}

In order to apply the Dirichlet boundary condition to compute the masses of oddballs, it requires the following condition on Eq. \eqref{anveccomp}
\begin{equation}
    A_{\nu} (z, x^{\mu})|_{z=z_{\rm max}}= 0 \Rightarrow J_{\nu}(m_{\nu, k} \,z)|_{z=z_{\rm max}} = 0\,,
\end{equation}
meaning that odd glueball masses will be given by the roots of the Bessel function. From the above equation, one has
\begin{equation}\label{massaD}
    m_{\nu, k}^D = \frac{\xi_{\nu, k} }{z_{\rm max}}\,, 
\end{equation}
where $\xi_{\nu, k}$ is the $k$ th zero of the Bessel function of order $\nu$. 
Due to a lack of experimental/theoretical data regarding higher radial excitation states for odd glueballs, we are going to focus only in the ground state and fix $k=1$. Then Eq. \eqref{massaD} becomes
\begin{equation}\label{massaDk1}
    m_{\nu, 1}^D = \frac{\xi_{\nu, 1} }{z_{\rm max}}\,.
\end{equation}

As we are interest higher odd spin glueballs, let us take a look at the Bessel function index $\nu$, in Eq. \eqref{anveccomp}. Such an index is related to the bulk mass $M_5$ by
\begin{equation}\label{index}
    \nu = \sqrt{M_5^2 +1}\,.
\end{equation}
Now, by plugging Eq. \eqref{hsi} in the above equation, one gets a relationship between the Bessel function index and the glueballs' angular momentum
\begin{equation}\label{indexl}
    \nu = \sqrt{(\Delta +\ell - p) (\Delta + \ell + p - 4)+1}\,.
\end{equation}
In particular, for the state $1^{--}$ one has $\ell=0$, $\Delta=6$, $p=1$ and then $\nu=4$. The IR cutoff ${z_{\rm max}}$  will be fixed by using the mass of this state, $m_{4, 1}^D$, as an input. 

At this moment, we can eliminate $z_{\rm max}$ by dividing an arbitrary odd spin state by the mass of the ground odd spin state $1^{--}$, in Eq. \eqref{massaDk1}, and get an expression to compute the masses of higher odd spin glueball states [$\ell\,\, \rm{(even)} \geq 2$], so that:
\begin{equation}\label{massageral}
    m_{4+\ell, 1}^D = \frac{\xi_{4+\ell, 1} }{\xi_{4, 1}}m_{4, 1}^D\,.
\end{equation}
Note that we will choose $m_{4, 1}^D = 3.02$ GeV as an input from \cite{Kaidalov:2005kz}. For this chosen input, one has ${z_{\rm max}}=2.51$ GeV$^{-1}$. 
This value for ${z_{\rm max}}$ was obtained from a chi-squared minimization procedure with the rms error given by
\begin{equation}\label{rms}
 \delta_{{\rm RMS}}= \sqrt{ \frac 1{N-N_p} \sum_{i=1}^N \left( \frac {\delta O_i}{O_i} \right)^2}  \times 100\,, 
\end{equation}
for the glueball masses present in  Table \ref{mgd} with Dirichlet boundary condition.  
Here $N$ is the number of measurements (glueball masses) and, $N_p=1$ is our only free parameter ($z_{\rm max}$).

Note that in the original hardwall model presented in Ref. \cite{Erlich:2005qh}, the $\rho$-meson mass was chosen to set the scale for the other particle masses. In that case, this is quite appropriate since they were considering three different meson families all with conformal dimension operator $\Delta=3$. In our case, the oddballs are characterized by $\Delta=6+\ell$. Thus it is natural for the present model to take the mass of the  oddball ground state $1^{--}$ to fix $z_{\rm max}$.

\bigskip 
\begin{table}
\centering
\begin{tabular}{|c|c|c|c|c|c|c|}
\hline
 &  \multicolumn{6}{c|}{Odd Glueball states $J^{PC}$}  \\  
\cline{2-7}
 & $1^{--}$ & $3^{--} $ & $5^{--}$ & $7^{--}$ & $9^{--}$ & $11^{--}$  \\
\hline \hline
Dirichlet b.c.                                   
&\, 3.02 \, &\, 3.95 \,&\, 4.87 \,& \, 5.76 \, &\, 6.45 \, &\, 7.52 \, \\ \hline 
\end{tabular}
\caption{ Odd spin glueball masses expressed in {\rm GeV} considering Dirichlet boundary condition, given by Eq. \eqref{massageral}.} 
\label{mgd}
\end{table}

Now, we are going to consider different sets of oddball states to construct possible odderon Regge trajectories. 
By considering the set $1^{--}, \cdots, 11^{--}$, and taking the masses in Table \ref{mgd}, one can construct the following Regge trajectory associated with the odderon:
\begin{equation}\label{rgd1}
    J_{{\rm Dir}}^{\{1-11\}} (m^2) = (0.21 \pm 0.01) m^2 - (0.35 \pm 0.48).
\end{equation}
Analogously, for the states $1^{--}, \cdots, 9^{--}$, one gets
\begin{equation}\label{rgd2}
    J_{{\rm Dir}}^{\{1-9\}} (m^2) = (0.24 \pm 0.01) m^2 - (0.95 \pm 0.24), 
\end{equation}
and for $3^{--}, \cdots, 11^{--}$, one finds
\begin{equation}\label{rgd3}
    J_{{\rm Dir}}^{\{3-11\}} (m^2) = (0.19 \pm 0.01) m^2 + (0.26 \pm 0.53).
\end{equation}

\begin{figure}[!ht]
\begin{center}
\includegraphics[scale=0.5]{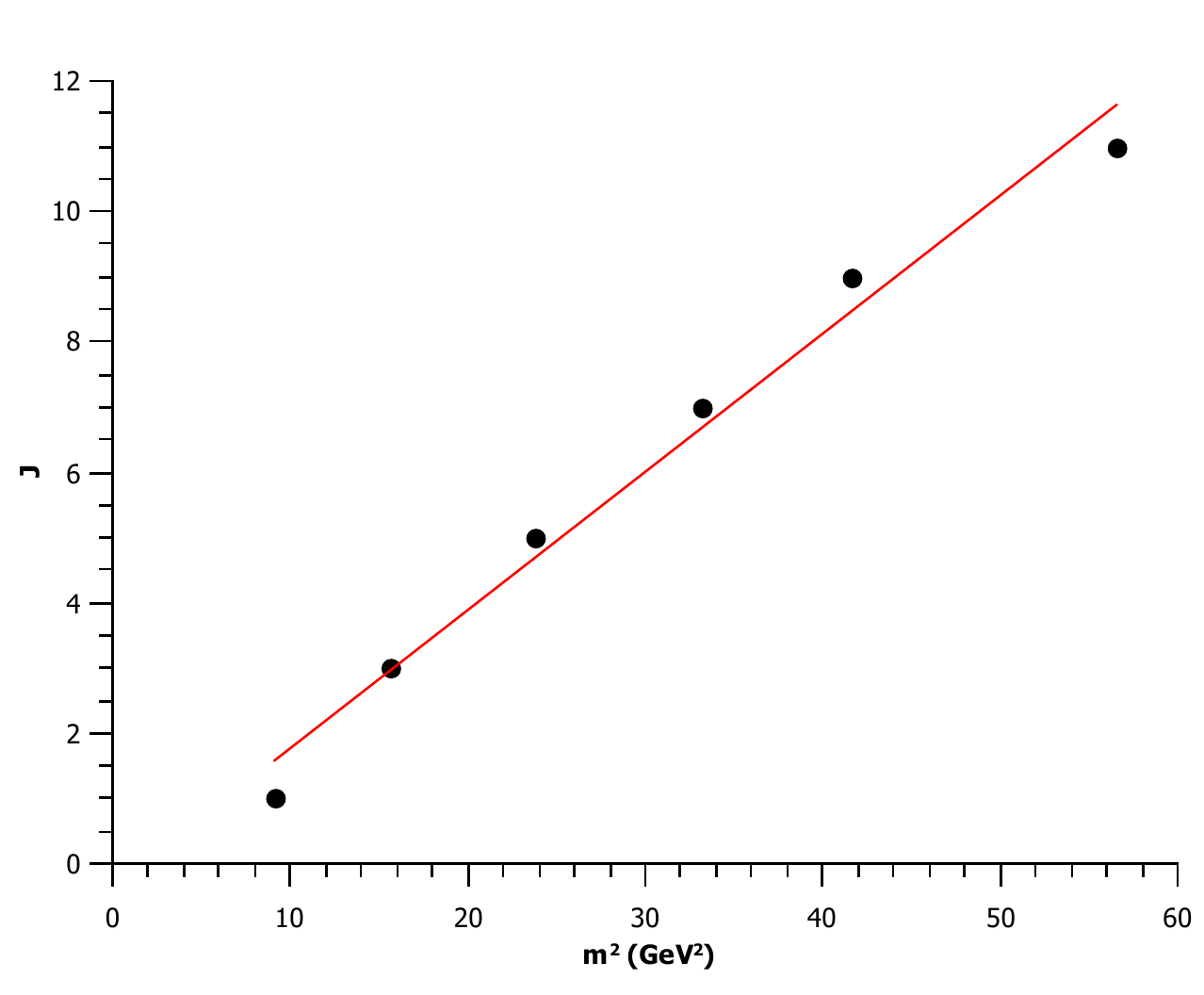}
\caption{Odderon Regge trajectory with Dirichlet boundary condition  corresponding to Eq. \eqref{rgd1}.}\label{dirbc1}
\end{center}
\end{figure}

\begin{figure}[!ht]
\begin{center}
\includegraphics[scale=0.5]{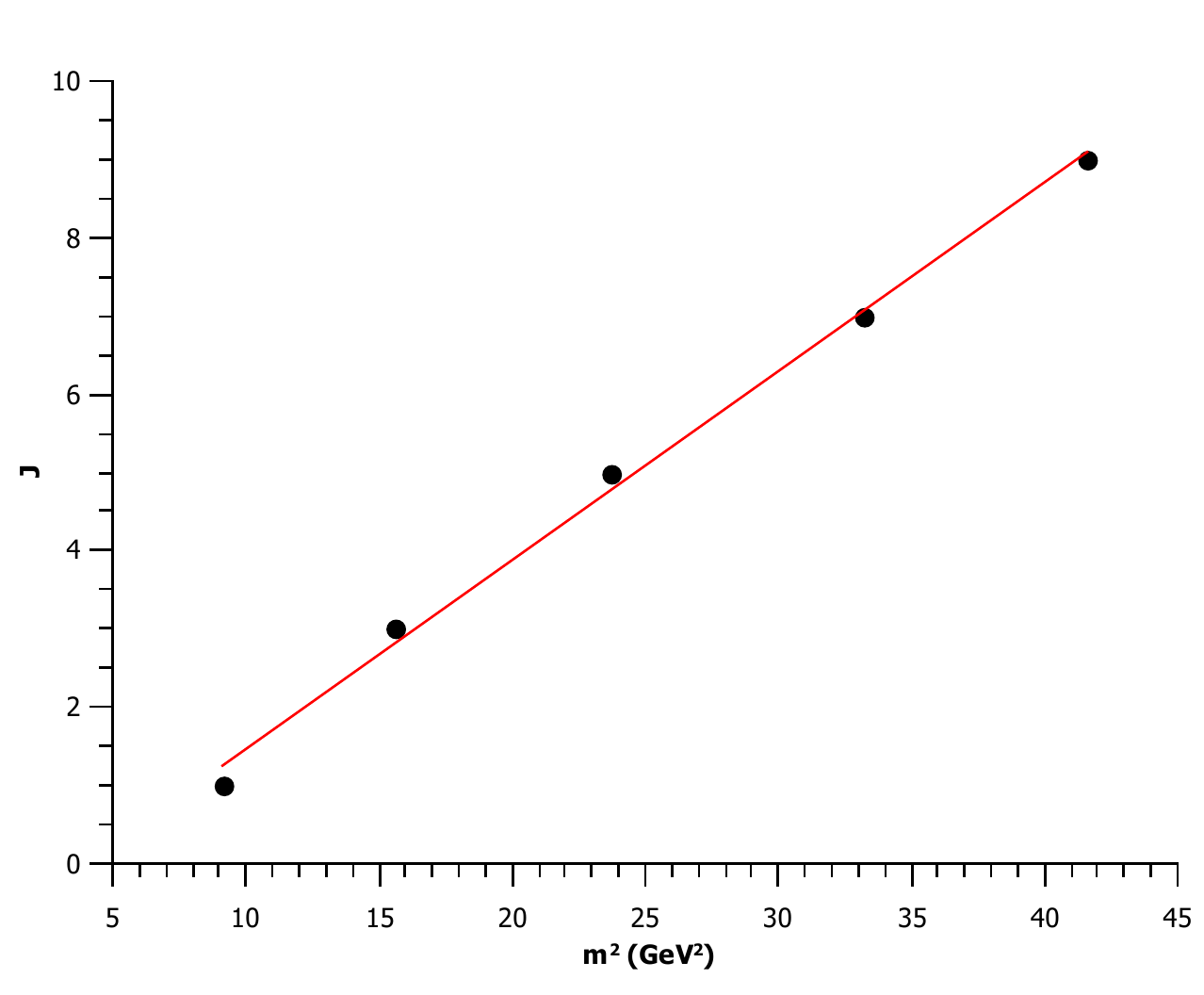}
\caption{Odderon Regge trajectory with Dirichlet boundary condition corresponding to Eq. \eqref{rgd2}.}\label{dirbc2}
\end{center}
\end{figure}

\begin{figure}[!ht]
\begin{center}
\includegraphics[scale=0.5]{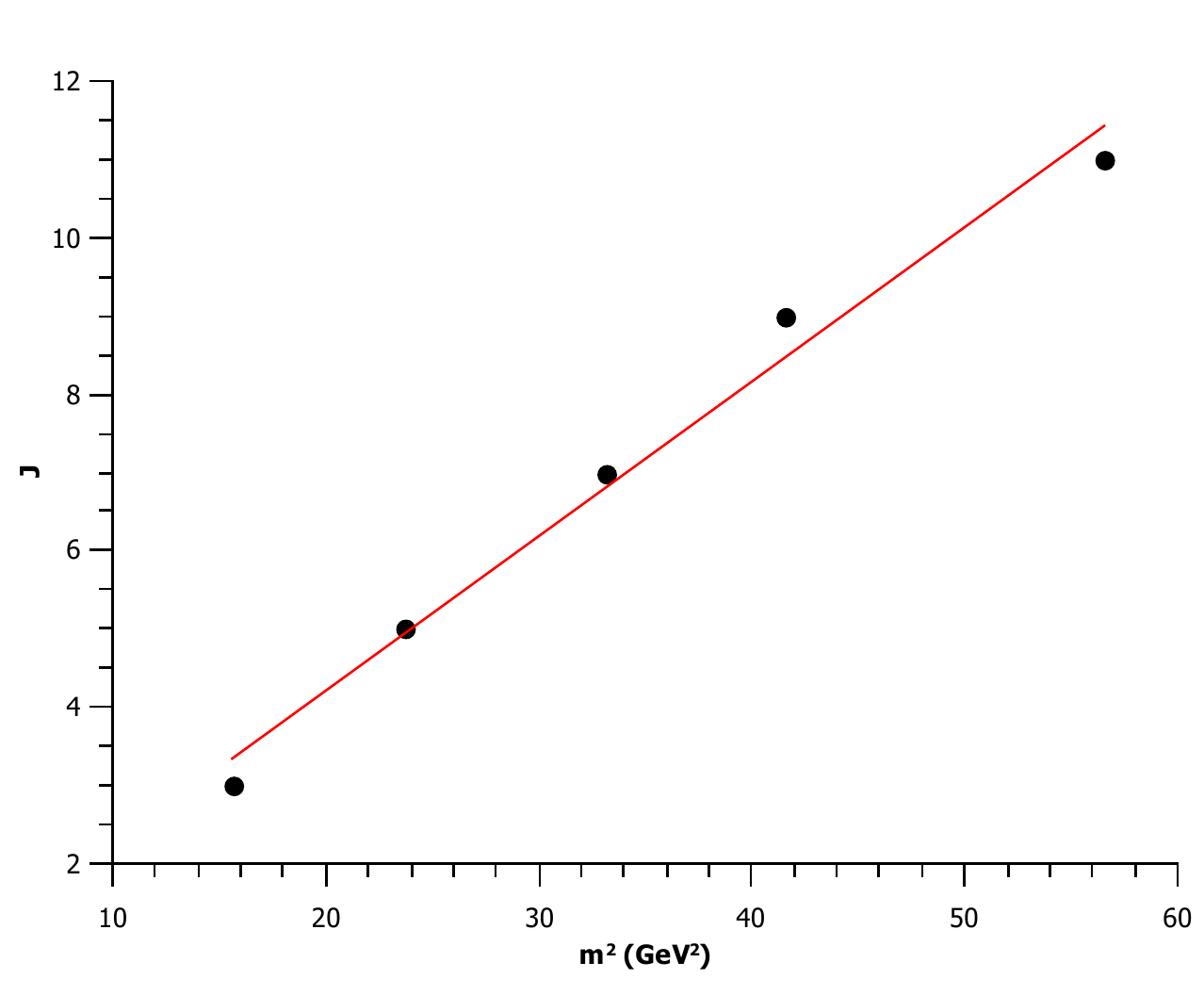}
\caption{Odderon Regge trajectory  with Dirichlet boundary condition corresponding to Eq. \eqref{rgd3}.}\label{dirbc3}
\end{center}
\end{figure}

It is worthwhile to mention that these Regge trajectories Eqs. \eqref{rgd1}-\eqref{rgd3} were obtained from a standard linear regression method by using the glueball masses from Table \ref{mgd}. The errors for the slope and intercept come from such an  analysis.    
These Regge trajectories are displayed in Figs. \ref{dirbc1}, \ref{dirbc2} and \ref{dirbc3}. 


\newpage 

\subsubsection{Neumann boundary condition}

For Neumann boundary condition on Eq. \eqref{anveccomp}, it requires
\begin{equation}
    \frac{d}{dz} A_{\nu} (z, x^{\mu})|_{z=z_{\rm max}}= 0 \Rightarrow  \frac{d}{dz} [z J_{\nu}(m_{\nu, k} \,z)]|_{z=z_{\rm max}} = 0\,.
\end{equation}
And then one gets
\begin{equation}\label{}
   J_{\nu}(m^N_{\nu, k} z_{\rm max}) + m^N_{\nu, k} z_{\rm max} \left [\frac{d}{dz}  J_{\nu }(m^N_{\nu, k} z_{\rm max})\right ] = 0\,.
\end{equation}
By using the following property:
\begin{equation}\label{}
    \frac{d}{dz}  J_{\alpha }(x) = J_{\alpha -1 }(x) - \frac{\alpha}{x} J_{\alpha}(x)\,,
\end{equation}
one has
\begin{equation}\label{massaN}
   m^N_{\nu, k} z_{\rm max} J_{\nu -1}(m^N_{\nu, k} z_{\rm max}) + (1- \nu) J_{\nu }(m^N_{\nu, k} z_{\rm max}) = 0\,,
\end{equation}
where the odd glueball mass computed in the hardwall model with Neumann boundary condition is given by
\begin{equation}\label{}
    m_{\nu, k}^N = \frac{\chi_{\nu, k} }{z_{\rm max}}\,.
\end{equation}
Here, we will fix ${z_{\rm max}}=1.89$ GeV$^{-1}$  with $m_{4, 1}^N = 3.02$ GeV, coming from \cite{Kaidalov:2005kz},  as an input. The value of ${z_{\rm max}}$ is determined by a chi-square minimization procedure analogous to the one done for Dirichlet boundary condition but now with the masses coming from Table \ref{t2}. 

To get higher odd spin glueball states we will proceed as done for Dirichlet boundary condition, and then we can rewrite Eq. \eqref{massaN} as
\begin{equation}\label{massan}
   \chi_{\nu+\ell, k} J_{\nu+\ell -1}(\chi_{\nu+\ell, k}) + (1- \nu + \ell) J_{\nu +\ell }(\chi_{\nu+\ell, k}) = 0\,,
\end{equation}
so that
\begin{equation}\label{massageralN}
    m_{\nu+\ell, k}^N = \frac{\chi_{\nu+\ell, k} }{z_{\rm max}}\,.
\end{equation}
As before, we just consider $k=1$ corresponding to nonexcited radial states. Then, from our model with Neumann boundary condition we get the set of masses, presented in Table \ref{t2}. 

\bigskip 
\begin{table}
\centering
\begin{tabular}{|c|c|c|c|c|c|c|}
\hline
 &  \multicolumn{6}{c|}{Odd Glueball states $J^{PC}$}  \\  
\cline{2-7}
 & $1^{--}$ & $3^{--} $ & $5^{--}$ & $7^{--}$ & $9^{--}$ & $11^{--}$  \\
\hline \hline
Neumann b.c.                                   
&\, 3.02 \, &\, 4.14 \,&\, 5.26 \,& \, 6.38 \, &\, 7.48 \, &\, 8.59 \, \\ \hline
\end{tabular}
\caption{ Odd spin glueball masses expressed in {\rm GeV}  considering Neumann boundary condition, given by Eq. \eqref{massageralN}.} 
\label{t2}
\end{table}

Considering different sets of oddball states to construct possible odderon Regge trajectories from our model with Neumann boundary condition, we get for  $1^{--}, \cdots, 11^{--}$,
\begin{equation}\label{rgn1}
    J_{{\rm Neu}}^{\{1-11\}} (m^2) = (0.16 \pm 0.01) m^2 + (0.33 \pm 0.45).
\end{equation}
In the same way, for $1^{--}, \cdots, 9^{--}$,
\begin{equation}\label{rgn2}
    J_{{\rm Neu}}^{\{1-9\}} (m^2) = (0.17 \pm 0.01) m^2 - (0.06\pm 0.41), 
\end{equation}
and for $1^{--}, \cdots, 5^{--}$,
\begin{equation}\label{rgn3}
    J_{{\rm Neu}}^{\{1-5\}} (m^2) = (0.22 \pm 0.02) m^2 - (0.83 \pm 0.30).
\end{equation}

\begin{figure}[!ht]
\begin{center}
\includegraphics[scale=0.5]{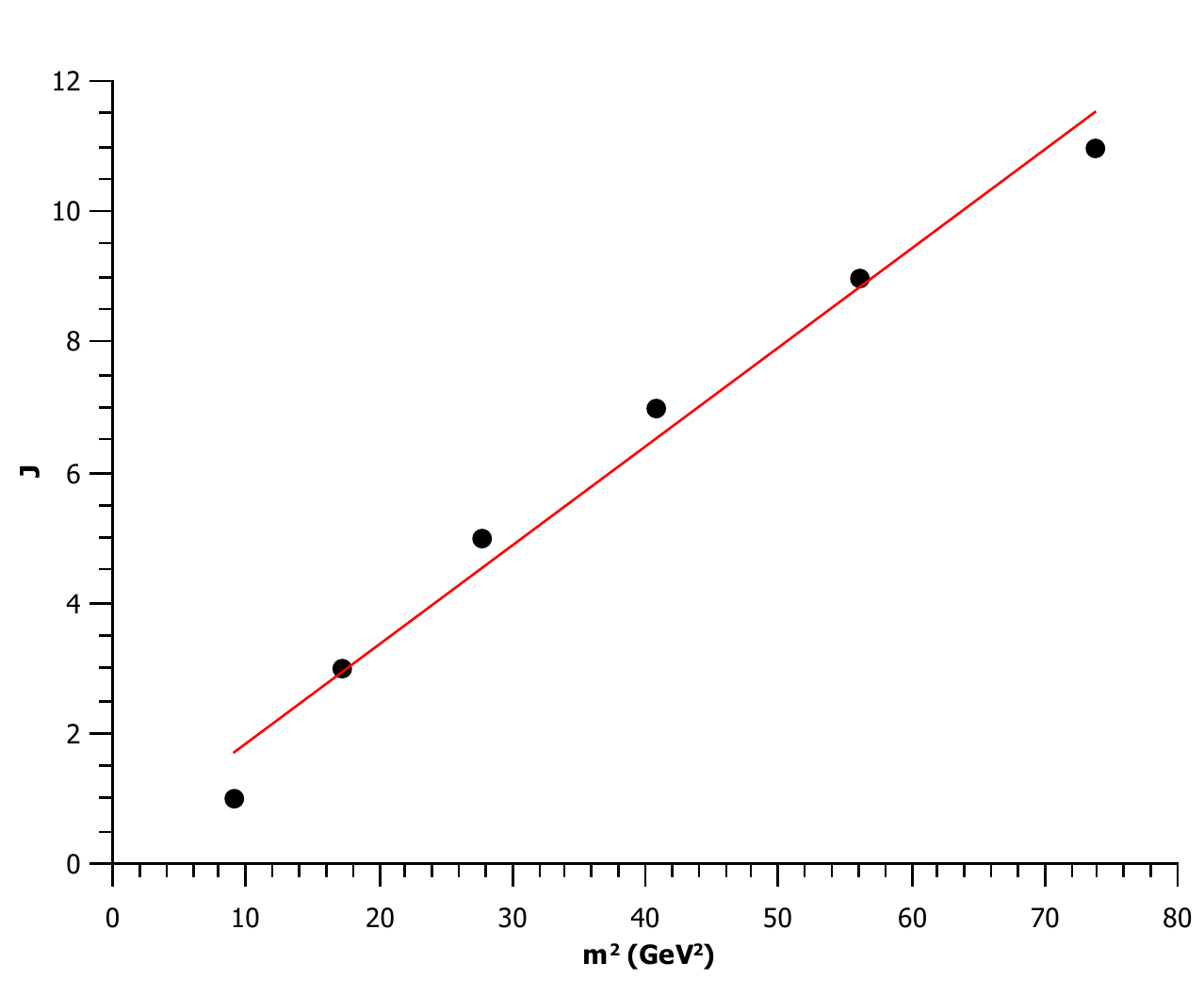}
\caption{Odderon Regge trajectory  with Neumann boundary condition corresponding to Eq. \eqref{rgn1}.}\label{neubc1}
\end{center}
\end{figure}

\begin{figure}[!ht]
\begin{center}
\includegraphics[scale=0.5]{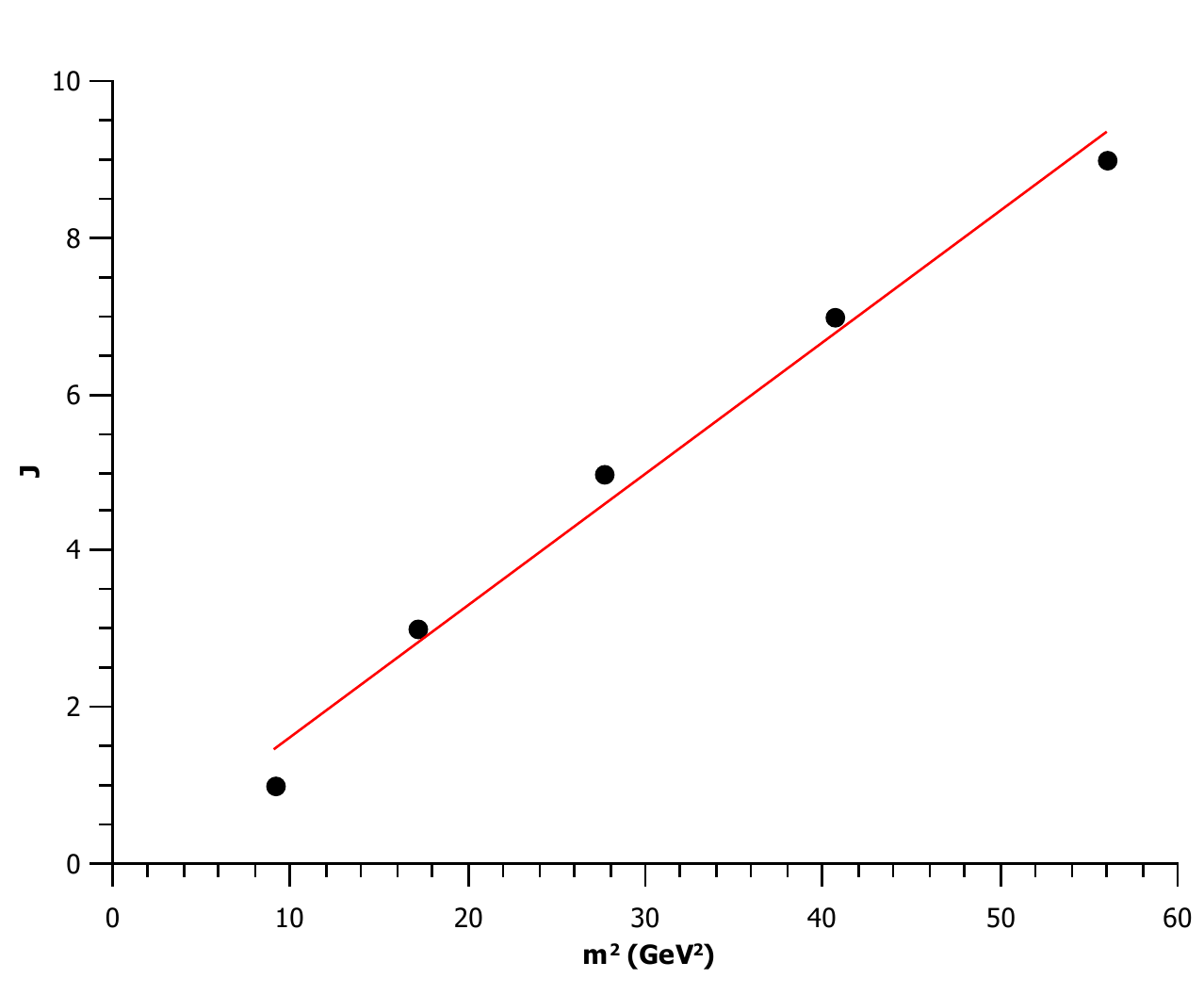}
\caption{Odderon Regge trajectory with Neumann boundary condition corresponding to Eq. \eqref{rgn2}.}\label{neubc2}
\end{center}
\end{figure}

\begin{figure}[!ht]
\begin{center}
\includegraphics[scale=0.5]{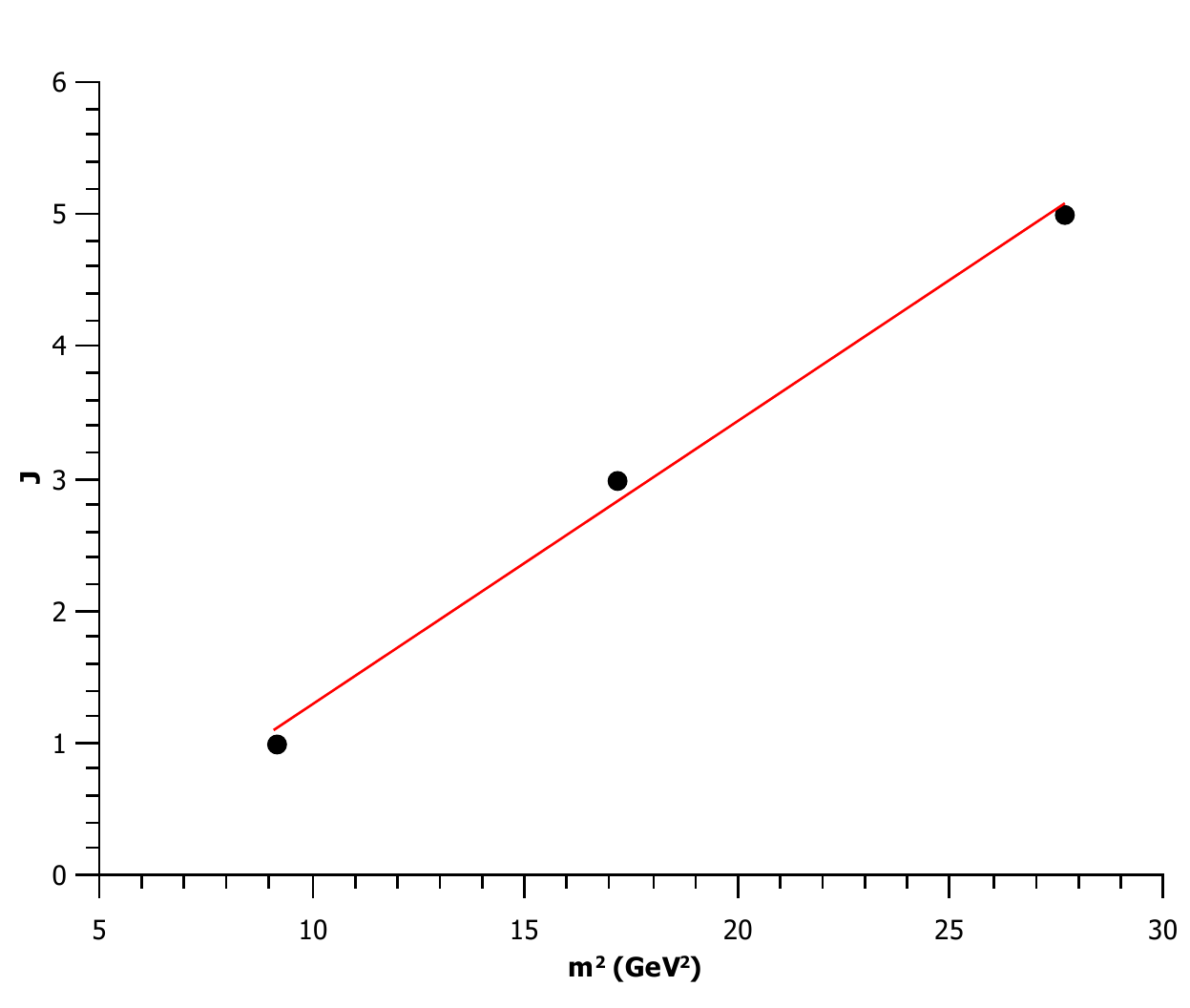}
\caption{Odderon Regge trajectory with Neumann boundary condition corresponding to Eq. \eqref{rgn3}.}\label{neubc3}
\end{center}
\end{figure}

Once again, these Regge trajectories Eqs. \eqref{rgn1}-\eqref{rgn3} were obtained from a standard linear regression method by using the glueball masses from Table \ref{t2}. The errors for the slope and intercept come from such an  analysis.
These Regge trajectories are displayed in Figs. \ref{neubc1}, \ref{neubc2}, and \ref{neubc3} .

In order to compare these results for the glueball masses, we are going to calculate the rms error with Eq. \eqref{rms}. 
Taking the values of the glueball masses from $1^{--}$ to $7^{--}$ of the vacuum correlator model in Ref. \cite{Kaidalov:2005kz} as our benchmarks, from \eqref{rms} with $N=4$, one finds that  $\delta_{\rm {RMS}} = 3.60 \%$ for the Dirichlet boundary condition from Table \ref{mgd} and  $\delta_{\rm {RMS}} = 5.61 \%$ for the Neumann boundary condition from Table \ref{t2}.  From this point of view the results for the hardwall model with twist-5 operator approach the Dirichlet boundary condition seems to work better.


\section{Conclusions}\label{conc}
In this section, we present our last comments on our work and present some interpretations on our achieved results.
Here, we have used the holographic hard wall model to compute the masses of odd spin glueball states from a twist-five operator approach as well as derive the corresponding Regge trajectories related to the Odderon with Dirichlet and Neumann boundary conditions. 

As the oddball ground state $1^{--}$ has spin 1 and the corresponding operator has conformal dimension $\Delta=6$, the twist of this state is $\tau=5$. In this sense, the twist-five operator approach seemed appropriated to deal with the odd glueball ground state. To implement it, we started with a massive gauge boson field living in the AdS$_5$  related to the vector glueball at the boundary theory.  
The higher spin oddballs $J^{--}=(1+\ell)^{--}$ (with even $\ell$) are then represented by operators with conformal dimension $\Delta+\ell=6+\ell$ so these states also have twist $\tau=5$. 

Note that one can wonder if it is possible to accommodate higher even spin glueball states in our  model. Note however that two possible even spin ground states $0^{++}$ and $2^{++}$ are a twist-4 or twist-2 objects. In our case we are dealing with just the twist-5 objects.

In order to compute the odd spin glueball masses, we had to introduce an  IR cutoff $z_{\rm max}$ by using the mass of ground state $1^{--}$ as an input. For our purposes, our input was taken from the vacuum correlator model as in Ref.  \cite{Kaidalov:2005kz}. Besides, in this reference one can also find values for higher odd spin glueballs masses as well as other Refs. mentioned in Table \ref{oddlit}. 

As one can see, the masses computed here for higher spin oddballs, by considering Dirichilet and Neumann boundary conditions (Tables \ref{mgd} and \ref{t2}, respectively) are fully compatible with most of the models presented in Table \ref{oddlit}. It is worth to mention that mass for the state $3^{--}$ computed in this work is also in agreement with the one obtained using a holographic QCD model as reported recently in Ref. \cite{Zhang:2021itx}. It is worth mentioning that in this work the results coming from the Dirichlet boundary condition seems to give better glueball masses than the Neumann one, taking as benchmarks the results from Ref. \cite{Kaidalov:2005kz}, since the respective rms errors are 3.60\% and 5.61\%, as discussed at the end of the previous section.

Another point of interest in this work is to derive, from the odd spin glueball masses, the Regge trajectories associate with the odderon. By taking a look at the masses in Table \ref{mgd}, within Dirichilet boundary condition, one can construct Regge trajectories for the odderon. For the oddballs considered in this work, from the ground state $1^{--}$ to the state $11^{--}$  and from the ground state $1^{--}$ to the state $9^{--}$, one can obtain the Regge trajectories presented in Eqs. \eqref{rgd1} and \eqref{rgd2}, respectively. These Regge trajectories are compatible with the one presented in Eq. \eqref{l1} within the RMB model of Ref. \cite{LlanesEstrada:2005jf}. On the other hand, if the one considers the set of the states $3^{--}, 5^{--}, 7^{--}, 9^{--}$ and $11^{--}$, the hardwall model used here, provides a Regge trajectory given by Eq. \eqref{rgd3} compatible with the one in Eq. \eqref{l2} within the NRCM, also in Ref. \cite{LlanesEstrada:2005jf}.

Regarding to the Neumann boundary condition, from Table \ref{t2}, one can also consider different sets of oddball states  and derive the corresponding Regge trajectories related to the odderon. The Regge trajectories presented in Eqs. \eqref{rgn1} and \eqref{rgn2}, considered the sets from the ground state $1^{--}$ to the state $11^{--}$ and to the state $9^{--}$  respectively, are compatible with the one presented in Eq. \eqref{l2} within the NRCM in Ref. \cite{LlanesEstrada:2005jf}. Nevertheless, the Regge trajectory in Eq. \eqref{rgn3}, considering the states $1^{--}, 3^{--}$ and $5^{--}$ is compatible with the one presented in Eq. \eqref{l1} within the RMB model of Ref. \cite{LlanesEstrada:2005jf}.

 One should notice that the values of odd spin glueball masses within Neumann boundary condition are greater than the ones coming from Dirichlet boundary condition. To build the Regge trajectory one has to choose a set of oddball  states. This feature implies that if one increases the number of elements in the chosen set, the slope of the Regge trajectory will decrease. This explains the difference between the slope and intercept of the Regge trajectories obtained in this work with both boundary conditions. 
 
 Even though the hardwall model may be the simplest among the AdS/QCD models  it provides good estimates of glueball masses despite the fact that the corresponding Regge trajectories are not intrinsically linear. Anyway, the hardwall model can provide approximate linear trajectories as the ones presented in this work compatible with other holographic and nonholographic approaches.
 In particular, one can note that the rms errors found here for glueballs are smaller than the corresponding ones for other hadrons as presented, for instance in Ref. \cite{Erlich:2005qh}.
 
 To conclude, we should keep in mind that although the oddballs discussed here are still lacking direct observation, the odderon itself was discovered experimentally  \cite{TOTEM:2017sdy, TOTEM:2020zzr}.  
 We hope that the oddball quest will come to a good end in future experiments.

\begin{acknowledgments}
J.P.M.G. is supported by Conselho Nacional de Desenvolvimento Científico e Tecnológico (CNPq) under Grant No. 151701/2020-2.
H.B.-F. is partially supported by Conselho Nacional de Desenvolvimento Científico e Tecnológico (CNPq) under Grant No. 311079/2019-9. This study was financed in part by the Coordenação de
Aperfeiçoamento de Pessoal de Nível Superior (CAPES), finance code 001. 
\end{acknowledgments}

{\it Note added}.—Recently Refs. \cite{Vega:2021yyj} and \cite{Rinaldi:2021xat} appeared on arXiv proposing a holographic study for glueballs at finite temperature, within softwall and hardwall models, respectively.

\end{document}